\documentclass[letterpaper, 10 pt, conference]{ieeeconf}  % Comment this line out
\IEEEoverridecommandlockouts                              % This command is only
\usepackage{graphics} % for pdf, bitmapped graphics files
\usepackage{times} % assumes new font selection scheme installed
\usepackage{amsmath, amsfonts} % assumes amsmath package installed
\usepackage{amssymb}  % assumes amsmath package installed
\usepackage{url}
\usepackage{cite}
\usepackage{hyperref}

\usepackage{algorithmic}
\usepackage{graphicx}
\usepackage{textcomp}
\usepackage{xcolor}

\title{\LARGE \bf
Parameter Estimation-Based Extended Observer for Linear Systems with Polynomial Overparameterization
}

\author{Anton Glushchenko, \textit{Member, IEEE} and Konstantin Lastochkin% <-this % stops a space
\thanks{A. Glushchenko is with V.A. Trapeznikov Institute of Control Sciences of Russian Academy of Sciences, Moscow, Russia
        {\tt\small aiglush@ipu.ru}}%
\thanks{K. Lastochkin is with V.A. Trapeznikov Institute of Control Sciences of Russian Academy of Sciences, Moscow, Russia 
        {\tt\small lastconst@ipu.ru}}%
}

\begin{document}

\maketitle
\thispagestyle{empty}
\pagestyle{empty}

\begin{abstract}
We consider a class of uncertain linear time-invariant overparametrized systems affected by bounded disturbances, which are described by a known exosystem with unknown initial conditions. For such systems an exponentially stable extended adaptive observer is proposed, which, unlike existing solutions, simultaneously: (\emph{i}) allows one to reconstruct original (physical) states of the system represented in arbitrarily chosen state-space form rather than virtual states of the observer canonical form; (\emph{ii}) ensures convergence of the state observation error to zero under weak requirement of the regressor finite excitation; (\emph{iii}) does not include Luenberger correction gain and forms states estimate using algebraic rather than differential equation; (\emph{iv}) additionally reconstructs the unmeasured external disturbance. The proposed solution is based on the new parametrizations to identify the observer parameters obtained with the help of the heterogeneous mappings and the dynamic regressor extension and mixing procedure. Illustrative simulations support obtained theoretical results.
\end{abstract}

\section{Introduction}
Adaptive observers are recurrent algorithms that simultaneously reconstruct unmeasured states and identify unknown parameters \cite{b1}. Nowadays various methods of such observers design have been proposed for linear SISO and MIMO systems \cite{b1, b2, b3, b4}, and also very important and interesting results have been obtained concerning state reconstruction of the nonlinear systems \cite{b5, b6} and ones with time-varying parameters \cite{b7, b8, b9}. In this study we return again to a deeply investigated problem of joint estimation of linear SISO systems state and parameters.

Retrospective analysis of existing solutions to the problem under consideration is given below. In \cite{b2, b3} adaptive observers for simultaneous estimation of states and unknown parameters of a linear SISO system have been proposed for the first time. G. Kreisselmeier \cite{b10} proposed a parameterization that, unlike in \cite{b2, b3}, allows one to completely separate the observer dynamics from the adaptive loop to make the design of suitable parameter adaptation schemes substantially simpler. R. Marino and P. Tomei in \cite{b11} generalized the result of \cite{b10} to linearizable nonlinear systems, and in \cite{b12} these results were additionally subjected to robustness stress tests. In \cite{b4}, motivated by \cite{b10, b11, b12}, an alternative observer with a simpler structure has been proposed. In \cite{b9} an exhaustive overview of some existing observer design algorithms for linearizable nonlinear systems with time-varying known parameters and time-invariant unknown parameters is given (see also the review \cite{b1}). The main drawbacks of all above-mentioned studies are two-fold:
\begin{enumerate}
\item[\textbf{D1)}] strict persistent excitation condition is required to guarantee convergence of the state observation and parameter errors;
\item[\textbf{D2)}] an appropriate state transformation is required to represent the system in a form without multiplications of unknown parameters and unmeasured states (the observer canonical form, for example).
\end{enumerate}

To overcome the first drawback (\textbf{D1}), two different approaches have been proposed recently. In \cite{b13}, a modified observer \cite{b10} with exponential convergence of state observation and parameter errors under the weaker regressor finite excitation requirement is proposed. In \cite{b14}, based on the results of \cite{b6, b7}, a parameter estimation-based observer is developed, which, unlike existing ones, \emph{i}) ensures the parameter error finite-time convergence when the regressor finite excitation condition is met, \emph{ii}) reconstructs states using algebraic rather than differential equation.

The second problem (\textbf{D2}) is more complex in comparison with the first one, since it does not allow one to apply the existing adaptive observers to reconstruct the physical states of dynamical systems. Instead, virtual states of appropriately chosen state space form, which includes only products between unknown parameters and measured functions (mappings for output and input signals), are recovered. At the same time, to solve many real-world control problems (\emph{e.g}. two-mass systems vibration suppression \cite{b15}) it is required to measure the physical states rather than the virtual ones.

In recent studies \cite{b16, b17}, two new adaptive observers of physical states of linear SISO systems with polynomial overparameterization have been proposed. The polynomial relation between the parameters of the observer canonical form and the original state space allows one to: (\emph{i}) parameterize and solve the problem of identification of parameters related to unmeasured states, (\emph{ii}) estimate the physical states of the system. The solution \cite{b16} is in line with the studies \cite{b2, b3, b9, b10, b13} and reconstructs the unmeasured state estimates using a differential equation, which is a copy of the system up to the Leuenberger corrective feedback. Observer in \cite{b17} is based on a parameter estimation-based observer design procedure \cite{b6, b7, b14} and, unlike \cite{b16}, forms states estimate using an algebraic equation and, unlike \cite{b14}, allows one to reconstruct the physical states of the system. Both proposed observers overcome \textbf{D1} and guarantee exponential convergence of the state observation error under weaker finite excitation condition. Comparing the observers from \cite{b16, b17} with each other, it should be noted that the solution in \cite{b17} does not use the Luenberger corrective feedback and, therefore, is not affected by the peaking phenomenon during transients (see Comparative Simulation in \cite{b16} and \cite{b17}).

In \cite{b18} the results of \cite{b16} are extended to the class of systems with external unmeasured disturbances generated by known exosystems with unknown initial conditions. In this study we extend the results of \cite{b17} in a similar way. The main advantages and salient features of the observer proposed in this research are summarized as follows:
\begin{enumerate}
\item[\textbf{F1)}] in comparison with \cite{b2, b3, b4, b8, b9, b10, b11, b12, b13, b14} physical states are reconstructed for a system represented in an arbitrary state space form rather than virtual states of the appropriately chosen state space form;
\item[\textbf{F2)}] the convergence of the unmeasured state estimates to their true values is guaranteed if the regressor finite excitation condition holds;
\item[\textbf{F3)}] in comparison with \cite{b16, b18}, the Luenberger corrective feedback is not applied, and state estimates are formed using an algebraic rather than a differential equation;
\item[\textbf{F4)}] in addition to the state and unknown parameter estimates, an unmeasured external disturbance generated by a known exosystem with unknown initial conditions is also reconstructed.
\end{enumerate}

The organization of this paper is as follows. Section II provides a rigorous problem statement. The design procedure for the proposed observer and its properties analysis are elucidated in Section III. Section IV presents some simulation results to validate the advantages of the proposed method in comparison with \cite{b16, b18}. The paper is wrapped up with conclusion and further research scope in Section V.

\textbf{Notation and Definitions.} Further the following notation is used: $\left| . \right|$ is the absolute value, $\left\| . \right\|$ is the suitable norm of $(.)$, ${I_{n \times n}}=I_{n}$ is an identity $n \times n$ matrix, ${0_{n \times n}}$ is a zero $n \times n$ matrix, $0_{n}$ stands for a zero vector of length $n$, ${\rm{det}}\{.\}$ stands for a matrix determinant, ${\rm{adj}}\{.\}$ represents an adjoint matrix. We also use the fact that for all (possibly singular) ${n \times n}$ matrices $M$ the following holds: ${\rm{adj}} \{M\} M  = {\rm{det}} \{M\}I_{n \times n}.$ For a mapping ${\mathcal F}{\rm{:\;}}{\mathbb{R}^n} \mapsto {\mathbb{R}^n}$  we denote its Jacobian by $\nabla_{x} {\mathcal F}\left( x \right) = \linebreak = {\textstyle{{\partial {\mathcal F}} \over {\partial x}}}\left( x \right)$. 

The below-given definitions are used throughout the paper.

{\it \bf Definition 1.} \emph{A mapping ${\cal F}{\rm{:\;}}{\mathbb{R}^{{n_x}}} \to {\mathbb{R}^{{n_{\cal F}} \times {m_{\cal F}}}}$ is heterogeneous of degree ${\ell _{\cal F}} \ge 1$ if there exists ${\Xi _{\cal F}}\left( {\omega} \right)=\linebreak = {\overline \Xi _{\cal F}}\left( {\omega} \right)\omega \left( t \right) \in {\mathbb{R}^{{\Delta _{\cal F}} \times {n_x}}}{\rm{,\;}}{\Pi _{\cal F}}\left( {\omega} \right) \in {\mathbb{R}^{{n_{\cal F}} \times {n_{\cal F}}}}$, and a mapping ${{\cal T}_{\cal F}}{\rm{:\;}}{\mathbb{R}^{{\Delta _{\cal F}}}} \to {\mathbb{R}^{{n_{\cal F}} \times {m_{\cal F}}}}$ such that for all $ \omega\left(t\right) \in \mathbb{R}$ and $x \in {\mathbb{R}^{{n_x}}}$ the following conditions hold:}
\begin{equation}\label{eq1}
\begin{array}{c}
{\Pi _{\cal F}}\left( {\omega} \right){\cal F}\left( x \right) = {{\cal T}_{\cal F}}\left( {{\Xi _{\cal F}}\left( {\omega} \right)x} \right){\rm{, }}\\
{\rm{det}}\left\{ {{\Pi _{\cal F}}\left( {\omega} \right)} \right\} \ge {\omega ^{{\ell _{_{\cal F}}}}}\left( t \right)\!{\rm{,\;}}\\{\Xi _{\cal F}}_{ij}\!\left( {\omega} \right) = {c_{ij}}{\omega ^{\ell_{ij}} }\left( t \right)\!{\rm{,\;}}{{\overline \Xi }_{{\cal F}ij}}\left( \omega  \right) = {c_{ij}}{\omega ^{{\ell _{ij}} - 1}}{\rm{,}}\\
{c_{ij}} \in \left\{ {0,{\rm{ 1}}} \right\}{\rm{,\;}}{\ell _\mathcal{F}} \geqslant 1,{\text{\;}}{\ell_{ij}}  \geqslant 1.
\end{array}
\end{equation}

For instance, for the mapping ${\cal F}\left( x \right) = {\rm{col}}\left\{ {{x_1}{x_2}{\rm{,\;}}{x_1}} \right\}$ with ${\Pi _{\cal F}}\left( \omega  \right) = {\rm{diag}}\left\{ {{\omega ^2}{\rm{,\;}}\omega } \right\}{\rm{,\;}}{\Xi _{\cal F}}\left( \omega  \right) = \omega{I_2},\;{\overline \Xi _{\cal F}}\left( \omega  \right) = {I_2}$ \linebreak we have that ${{\cal T}_{\cal F}}\left( {{\Xi _{\cal F}}\left( \omega  \right)x} \right) = {{\cal T}_{\cal F}}\left( {{{\overline \Xi }_{\cal F}}\left( \omega  \right)\omega x} \right) =\linebreak={\rm{col}}\left\{ {\omega {x_1}\omega {x_2}{\rm{,\;}}\omega {x_1}} \right\}$.

Definition 1 is useful, for example, for the following task. Let us suppose that we need to measure (compute) the multiplication ${\Pi _{\cal F}}\left( \omega  \right){\cal F}\left( x \right)$, where ${\Pi _{\cal F}}\left( \omega  \right)$ is a measurable/computable signal and ${\cal F}\left( x \right)$ is unknown. Then, if we have a measurable signal ${\cal Y}\left( t \right) = \omega \left( t \right)x$  with $\omega \left( t \right) \in \mathbb{R}$, then, owing to \eqref{eq1}, the solution to Task 1 is a measurable signal ${{\cal T}_{\cal F}}\left( {{{\overline \Xi }_{\cal F}}\left( \omega  \right){\cal Y}} \right)$. For instance, for the example under consideration ${{\cal T}_{\cal F}}\left( {{{\overline \Xi }_{\cal F}}\left( \omega  \right){\cal Y}} \right) = {\rm{col}}\left\{ {{{\cal Y}_1}{{\cal Y}_2}{\rm{,\;}}{{\cal Y}_1}} \right\} = \linebreak = {\Pi _{\cal F}}\left( \omega  \right){\cal F}\left( x \right)$. Moreover, from such a regression ${\cal F}\left( x \right)$ can be identified directly without estimation of $x$ and application of the substitution ${\cal F}\left(\hat x \right)$.

{\it \bf Definition 2.} \emph{The regressor $\varphi \left( t \right) \in {\mathbb{R}^n}$ is finitely exciting $(\varphi \in {\rm{FE}})$  over $\left[ {t_r^ + {\rm{,\;}}{t_e}} \right]$  if there exists $t_r^ +  \ge 0$, ${t_e} > t_r^ +$ and $\alpha$ such that the following inequality holds:}
\begin{equation}\label{eq2}
\int\limits_{t_r^ + }^{{t_e}} {\varphi \left( \tau  \right){\varphi ^{\rm{T}}}\left( \tau  \right)d} \tau  \ge \alpha {I_n}{\rm{,}}
\end{equation}
\emph{where $\alpha > 0$ is the excitation level, $I_{n}$ is an identity matrix.}

The inequality \eqref{eq2} is a necessary and sufficient condition for identifiability of unknown parameters $x \in {\mathbb{R}^n}$ of a regression equation $y\left( t \right) = {\varphi ^{\rm{T}}}\left( t \right)x$ \cite{b100}.

\section{Problem Statement}

Uncertain linear time-invariant overparametrized systems affected by bounded external disturbances are considered\footnote{Dependencies from $\theta$ and $t$ can be further suppressed for brevity.}:
\begin{equation}\label{eq3}
\begin{array}{l}
\dot x\left( t \right) = A\left( \theta  \right)x\left( t \right) + B\left( \theta  \right)u\left( t \right) + D\left( \theta  \right)\delta \left( t \right){\rm{,}}\\
y\left( t \right) = {C^{\rm{T}}}x\left( t \right){\rm{,\;}}x\left( {{t_0}} \right) = {x_0}{\rm{,}}
\end{array}
\end{equation}
where $x\left( t \right) \in {\mathbb{R}^n}$ is the original (physical) system states with unknown initial conditions ${x_0}$, $\delta \left( t \right) \in \mathbb{R}$ is a bounded disturbance, $A{\rm{:\;}}{\mathbb{R}^{{n_\theta }}} \!\to\! {\mathbb{R}^{n \times n}}{\rm{,}}$ $B{\rm{:\;}}{\mathbb{R}^{{n_\theta }}} \!\to\! {\mathbb{R}^n}{\rm{,\;}}$ $D{\rm{:\;}}{\mathbb{R}^{{n_\theta }}} \!\to\! {\mathbb{R}^n}$ denote known mappings with unknown parameter $\theta  \in {\mathbb{R}^{{n_\theta }}}$, $C{\text{\;:=\;}}C(\theta)\in {\mathbb{R}^n}$ stands for a known constant vector or a mapping that depends only on $\theta$. The pair $\left( {{C^{\rm{T}}}{\rm{,\;}}A\left( \theta  \right)} \right)$ is completely observable with $\theta \!\in\! D_{\theta}$ and only control $u\left( t \right) \!\in\! \mathbb{R}$ and output $y\left( t \right) \in \mathbb{R}$ signals are measurable.

Considering the control signal, disturbances and structure of the system, the following assumptions are adopted.

{\it \bf Assumption 1.} \emph{For all $t \ge {t_0}$ the control signal $u\left( t \right)$ ensures existence and boundedness of trajectories of \eqref{eq3}.}

{\it \bf Assumption 2.} \emph{The disturbance $\delta \left( t \right)$ is bounded, continuous and generated by a time-invariant exosystem:
\begin{equation}\label{eq4}
\begin{array}{l}
{{\dot x}_\delta }\left( t \right) = {{\cal A}_\delta }{x_\delta }\left( t \right){\rm{,\;}}{x_\delta }\left( {{t_0}} \right) = {x_{\delta 0}}{\rm{,}}\\
\delta \left( t \right) = h_\delta ^{\rm{T}}{x_\delta }\left( t \right){\rm{,}}
\end{array}
\end{equation}
where ${x_\delta }\left( t \right) \in {\mathbb{R}^{{n_\delta }}}$ are exosystem states with unknown initial conditions ${x_{\delta 0}}$, ${h_\delta } \in {\mathbb{R}^{{n_\delta }}}{\rm{,\;}}{{\cal A}_\delta } \in {\mathbb{R}^{{n_\delta } \times {n_\delta }}}$ are known vector and matrix such that the pair $\left( {h_\delta ^{\rm{T}}{\rm{,\;}}{{\cal A}_\delta }} \right)$ is observable.}

{\it \bf Assumption 3.} \emph{The parameters $\theta$ are globally structurally identifiable, i.e. for almost any $\underline\theta\in D_{\theta}$, the following hold}
\begin{gather*}
    y\left(t,\;\underline{\theta},\; u\right)=y\left(t,\;\theta,\;u\right),\;\forall t \ge t_{0},\;\forall u \in \mathbb{R}
\end{gather*}
{\emph{only for $\underline{\theta}=\theta.$}}

The goal is to design an adaptive observer, which ensures that the following equalities hold:
\begin{equation}\label{eq5}
\mathop {{\rm{lim}}}\limits_{t \to \infty } \left\| {\tilde x\left( t \right)} \right\| = 0{\rm{\;}}\left( {\exp } \right){\rm{,\;}}\mathop {{\rm{lim}}}\limits_{t \to \infty } \left\| {\tilde \delta \left( t \right)} \right\| = 0{\rm{\;}}\left( {\exp } \right){\rm{,}}
\end{equation}
where $\tilde x\left( t \right) = \hat x\left( t \right) - x\left( t \right)$ is a state observation error of the system \eqref{eq3}, $\tilde \delta \left( t \right) = \hat \delta \left( t \right) - \delta \left( t \right)$ is a disturbance observation error, $\left(\rm{exp}\right)$ is an abbreviation for exponential rate of convergence. 

\section{Main Result}

In accordance with the results from \cite{b2}, for each completely observable linear system \eqref{eq3} for all $\theta \in D_{\theta}$ there exist nonsingular matrices:
\begin{gather*}
{\small{
\begin{array}{c}
{T_{I}}\left( \theta  \right) \!=\! {\begin{bmatrix}
{{A^{n - 1}}\left( \theta  \right){{\cal O}_n}\left( \theta  \right)}&{{A^{n - 2}}\left( \theta  \right){{\cal O}_n}\left( \theta  \right)}& \cdots &{{{\cal O}_n}\left( \theta  \right)}
\end{bmatrix}}{\rm{,}}\\
{{\cal O}_n}\left( \theta  \right) = {\cal O}\left( \theta  \right){{\begin{bmatrix}
{{0_{1 \times \left( {n - 1} \right)}}}&1
\end{bmatrix}}^{\rm{T}}}{\rm{,}}\\
{{\cal O}^{ - 1}}\!\left( \theta  \right) \!=\! {{\begin{bmatrix}
{C}&{{{\left( {A\left( \theta  \right)} \right)}^{\rm{T}}}C}& \!\cdots\! &{{{\left( {{A^{n - 1}}\left( \theta  \right)} \right)}^{\rm{T}}}C}
\end{bmatrix}}^{\rm{T}}}{\rm{,}}
\end{array}}}
\end{gather*}
which define the similarity transformation $\xi \left( t \right) = T\left( \theta  \right)x\left( t \right)$ to rewrite the system \eqref{eq3} in the observer canonical form:
\begin{equation}\label{eq7}
\dot \xi \left( t \right)\! =\! {A_0}\xi \left( t \right) + {\psi _a}\left( \theta  \right)y\left( t \right) + {\psi _b}\left( \theta  \right)u\left( t \right)+{\psi _d}\left( \theta  \right)\delta \left( t \right){\rm{,}}
\end{equation}
\begin{equation}\label{eq8}
y\left( t \right) = C_0^{\rm{T}}\xi \left( t \right){\rm{,\;}}\xi \left( {{t_0}} \right) = {\xi _0}\left( \theta  \right) = T\left( \theta  \right){x_0}{\rm{,}}
\end{equation}
where
\begin{gather*}
\begin{array}{c}
{\psi _a}\left( \theta  \right) = T\left( \theta  \right)A\left( \theta  \right){T^{ - 1}}\left( \theta  \right){C_0}{\rm{,\;}}{\psi _b}\left( \theta  \right) = T\left( \theta  \right)B\left( \theta  \right){\rm{,}}\\
{\psi _d}\left( \theta  \right) = T\left( \theta  \right)D\left( \theta  \right){\rm{,}}\\
{A_0} = {\begin{bmatrix}
{{0_n}}&{\begin{array}{*{20}{c}}
{{I_{n - 1}}}\\
{{0_{1 \times \left( {n - 1} \right)}}}
\end{array}}
\end{bmatrix}}{\rm{,\;}}\begin{array}{*{20}{c}}
{C_0^{\rm{T}} = {C^{\rm{T}}}{T^{ - 1}}\left( \theta  \right) = }\\
{ = {\begin{bmatrix}
1&{0_{n - 1}^{\rm{T}}}
\end{bmatrix}} }
\end{array}{\rm{,}}
\end{array}
\end{gather*}
${T_{I}}\left( \theta  \right){\rm{:}}=T^{-1}\left( \theta  \right)$, ${{\cal O}_n}$ is the $n^{th}$ column of the matrix that is inverse to ${{\cal O}^{ - 1}}\left( \theta  \right)$, $\xi \left( t \right) \in {\mathbb{R}^n}$ denotes state vector of the observer canonical form with unknown initial conditions ${\xi _0}$, the vector ${C_0} \in {\mathbb{R}^n}$ and mappings ${\psi_a}{\rm{,\;}}{\psi _b}{\rm{,\;}}{\psi _d}{\rm{:\;}}{\mathbb{R}^{{n_\theta }}} \to {\mathbb{R}^n}$ are known.

The similarity transformation $\xi \left( t \right) = T\left( \theta  \right)x\left( t \right)$ and representation \eqref{eq7}, \eqref{eq8} motivate to reconstruct the unmeasured states $x\left( t \right)$ in the following way:
\begin{equation}\label{eq9}
\hat x\left( t \right) = {\hat T_I}\left( t \right)\hat \xi \left( t \right){\rm{,}}
\end{equation}
where $\hat \xi \left( t \right)$ is the estimate of observer canonical form \eqref{eq7} states, ${\hat T_I}\left( t \right)$ stands for the estimate of the matrix ${T_I}\left( \theta  \right)$.

At the same time, the solution of the set of equations \eqref{eq4} is written as:
\begin{equation}\label{eq10}
\begin{array}{l}
{{\dot \Phi }_\delta }\left( t \right) = {{\cal A}_\delta }{\Phi _\delta }\left( t \right){\rm{,\;}}{\Phi _\delta }\left( {{t_0}} \right) = {I_{{n_\delta }}}{\rm{,}}\\
{x_\delta }\left( t \right) = {\Phi _\delta }\left( t \right){x_{\delta 0}}{\rm{,}}\\
\delta \left( t \right) = h_\delta ^{\rm{T}}{\Phi _\delta }\left( t \right){x_{\delta 0}}{\rm{,}}
\end{array}
\end{equation}
which motivate to estimate the external disturbance $\delta \left( t \right)$ as:
\begin{equation}\label{eq11}
\hat \delta \left( t \right) = h_\delta ^{\rm{T}}{\Phi _\delta }\left( t \right){\hat x_{\delta 0}}\left( t \right){\rm,}
\end{equation}
where ${\Phi }_\delta\left( t \right)\in\mathbb{R}^{n_{\delta}}$ is the fundamental solution of \eqref{eq4}.

According to equations \eqref{eq9} and \eqref{eq11}, the problem \eqref{eq5} of unmeasured signals estimation is transformed into the problem of identification of unknown parameters ${x_{\delta 0}}{\rm{,\;}}{T_I}\left( \theta  \right)$ and estimation of unmeasured states $\xi \left( t \right)$ of the observer canonical form \eqref{eq7}. According to \cite{b6, b7, b14}, the problem of $\xi \left( t \right)$ estimation can be reduced to the one of parameter identification. Thus, using the results \cite{b14, b18, b19}, the following parameterizations are obtained for the unknown parameters $\eta \left( \theta  \right) = {\rm{col}}\left\{ {{\psi _a}\left( \theta  \right){\rm{,\;}}{\psi _b}\left( \theta  \right)} \right\}$ and states $\xi \left( t \right)$.

{\bf{Lemma 1.}} \emph{Let ${t_\epsilon} > {t_0}$ be a sufficiently large predefined time instant, then for all $t \ge {t_\epsilon}$ the unknown parameters $\eta \left( \theta  \right)$ and unmeasured states $\xi \left( t \right)$ satisfy the following regression models:}
\begin{equation}\label{eq12}
\begin{array}{c}
{\cal Y}\left( t \right) = \Delta \left( t \right)\eta \left( \theta  \right),
\end{array}
\end{equation}
\vspace{-25pt}
\begin{gather*}
 {\cal Y}\left( t \right) = k\left( t \right) \cdot {\rm{adj}}\left\{ {\varphi \left( t \right)} \right\}q\left( t \right){\rm{,\;}}\Delta \left( t \right) = k\left( t \right) \cdot {\rm{det}}\left\{ {\varphi \left( t \right)} \right\},   
\end{gather*}
\vspace{-15pt}
\begin{equation}\label{eq13}
\begin{array}{c}
\xi \left( t \right) = z\left( t \right) + {R^{\rm{T}}}\left( t \right)\kappa \left( \theta  \right){\rm{,}}\\
{\rm{ }}\kappa \left( \theta  \right) = {{\begin{bmatrix}
{\psi _a^{\rm{T}}\left( \theta  \right)}&{\psi _b^{\rm{T}}\left( \theta  \right)}&{\psi _d^{\rm{T}}\left( \theta  \right)}
\end{bmatrix}}^{\rm{T}}}{\rm{,}}\\
{R^{\rm{T}}}\left( t \right) = {\begin{bmatrix}
{\Omega \left( t \right)}&{P\left( t \right)}&{U\left( t \right)}
\end{bmatrix}}{\rm{,}}
\end{array}
\end{equation}
\emph{where}
\begin{gather*}
\begin{array}{c}
    q\left( t \right) = \int\limits_{{t_\epsilon}}^t {{e^{-\sigma(\tau-t_{\epsilon})}}{{\overline \varphi }_f}\left( \tau  \right)\!\left(\!\! \begin{array}{c}
\overline q\left( \tau  \right) - {k_1}{{\overline q}_f}\left( \tau  \!\!\right) - \\
 - {\beta ^{\rm{T}}}\left( {{F_f}\left( \tau  \right) + l{y_f}\left( \tau  \right)} \right)
\end{array} \right)d\tau } {\rm{,}}\\
{\rm{ }}q\left( {{t_\epsilon}} \right) = {0_{{\rm{2}}n}},\\
\end{array}
\end{gather*}
\vspace{-10pt}
\begin{equation}\label{eq14}
\begin{array}{c}
\varphi \left( t \right) = \int\limits_{{t_\epsilon}}^t {{e^{ -\sigma(\tau-t_{\epsilon}) }}{{\overline \varphi }_f}\left( \tau  \right)\overline \varphi _f^{\rm{T}}\left( \tau  \right)d\tau } {\rm{,\;}}\\
\varphi \left( {{t_\epsilon}} \right) = {0_{{\rm{2}}n \times {\rm{2}}n}},
\end{array}
\end{equation}
\begin{equation}\label{eq15}
\begin{array}{l}
{{\dot {\overline q}}_f}\left( t \right) =  - {k_1}{{\overline q}_f}\left( t \right) + \overline q\left( t \right){\rm{,\;}}{{\overline q}_f}\left( {{t_0}} \right) = 0,\\
{{\dot {\overline \varphi} }_f}\left( t \right) =  - {k_1}{{\overline \varphi }_f}\left( t \right) + \overline \varphi \left( t \right){\rm{,\;}}{{\overline \varphi }_f}\left( {{t_0}} \right) = {0_{{\rm{2}}n}},\\
{{\dot F}_f}\left( t \right) =  - {k_1}{F_f}\left( t \right) + F\left( t \right){\rm{,\;}}{F_f}\left( {{t_0}} \right) = {0_{{n_\delta }}},\\
{{\dot y}_f}\left( t \right) =  - {k_1}{y_f}\left( t \right) + y\left( t \right){\rm{,\;}}{y_f}\left( {{t_0}} \right) = 0,
\end{array}
\end{equation}
\begin{equation}\label{eq16}
\begin{array}{c}
\overline q\left( t \right) = y\left( t \right) - C_0^{\rm{T}}z{\rm{,\;}}\overline \varphi \left( t \right) = {\begin{bmatrix}
{{{\dot \Omega }^{\rm{T}}}{C_0} + {N^{\rm{T}}}\beta }\\
{{{\dot P}^{\rm{T}}}{C_0} + {H^{\rm{T}}}\beta }
\end{bmatrix}}{\rm{,}}\\
\dot z\left( t \right) = {A_K}z\left( t \right) + Ky\left( t \right){\rm{,\;}}z\left( {{t_0}} \right) = {0_n}{\rm{,}}\\
\dot \Omega \left( t \right) = {A_K}\Omega \left( t \right) + {I_n}y\left( t \right){\rm{,\;}}\Omega \left( {{t_0}} \right) = {0_{n \times n}}{\rm{,}}\\
\dot P\left( t \right) = {A_K}P\left( t \right) + {I_n}u\left( t \right){\rm{,\;}}P\left( {{t_0}} \right) = {0_{n \times n}}{\rm{,}}\\
\end{array}
\end{equation}
\vspace{-15pt}
\begin{gather*}
\begin{array}{c}
    \dot U\left( t \right) = {A_K}U\left( t \right) + {I_n}\delta \left( t \right){\rm{,\;}}U\left( {{t_0}} \right) = {0_{n \times n}}{\rm{,}}\\
\dot F\left( t \right) = GF\left( t \right) + Gly\left( t \right) - lC_0^{\rm{T}}\dot z\left( t \right){\rm{,\;}}F\left( {{t_0}} \right) = {0_{{n_\delta }}}{\rm{,}}\\
\dot H\left( t \right) = GH\left( t \right) - lC_0^{\rm{T}}\dot P\left( t \right){\rm{,\;}}H\left( {{t_0}} \right) = {0_{{n_\delta } \times n}}{\rm{,}}\\
\dot N\left( t \right) = GN\left( t \right) - lC_0^{\rm{T}}\dot \Omega \left( t \right){\rm{,\;}}N\left( {{t_0}} \right) = {0_{{n_\delta } \times n}}{\rm{,}}
\end{array}
\end{gather*}
\emph{and, if $\overline \varphi \in {\rm{FE}}$ over $\left[ {{t_\epsilon}{\rm{,\;}}{t_e}} \right]$, then for all $t \ge {t_e}$ it holds that ${\Delta _{{\rm{max}}}} \ge \Delta \left( t \right) \ge {\Delta _{{\rm{min}}}} > 0$.}

\emph{Here $k\left( t \right) \!>\! k_{min} \!>\! 0$ is a time-varying (or time-invariant) amplifier, ${k_1} > 0$, $\sigma  > 0$ are filters time constants, ${A_K} = \linebreak ={A_0} - KC_0^{\rm{T}}{\rm{,\;}}G$ stand for stable matrices of appropriate dimensions, $l \in {\mathbb{R}^{{n_\delta }}}$  denotes a vector such that the pair $\left( {G{\rm{,\;}}l} \right)$ is controllable, and $G$ is chosen so as to satisfy the condition $\sigma \left\{ {{{\cal A}_\delta }} \right\} \cap \sigma \left\{ G \right\} = 0$, $\beta  \in {\mathbb{R}^{{n_\delta }}}$ is a solution of the following set of equations:}
\begin{gather*}
\begin{array}{l}
{M_\delta }{{\cal A}_\delta } - G{M_\delta } = l\overline h_\delta ^{\rm{T}}{\rm{,\;}}\overline h_\delta ^{\rm{T}} = h_\delta ^{\rm{T}}{{\cal A}_\delta }{\rm{,}}\\
\beta  = \overline h_\delta ^{\rm{T}}M_\delta ^{ - 1}.
\end{array}
\end{gather*}

\emph{Proof of Lemma is given in Supplementary material \cite{b20}.}

~

Following parametrizations \eqref{eq9}, \eqref{eq11}, \eqref{eq13}, in order to estimate the states $x\left( t \right)$, it is sufficient to obtain the estimates of $\kappa \left( \theta  \right){\rm{,\;}}{T_I}\left( \theta  \right)$ and ${x_{\delta 0}}.$ However, the regression equation \eqref{eq12} allows one to find only the parameters $\eta \left( \theta  \right)$. Therefore, it is required to: \emph{a}) parametrize an equation with respect to (w.r.t.) ${x_{\delta 0}}$ and \emph{b}) transform \eqref{eq12} into the regression equations w.r.t. the parameters $\kappa \left( \theta  \right){\rm{,\;}}{T_I}\left( \theta  \right)$.

As the parameters  $\theta$ are globally structurally identifiable thanks to the Assumption 3, then, in accordance with \cite{b21}, the following existence condition of inverse function  ${\cal F} {\rm{:\;}}{\mathbb{R}^{{n_\theta }}} \to {\mathbb{R}^{{n_\theta }}},\; \psi_{ab}\mapsto\theta$ is met:
\begin{equation}\label{eq17}
\begin{array}{c}
{{\rm{det}} ^2}\left\{ {{\nabla _\theta }{\psi _{ab}}\left( \theta  \right)} \right\} > 0,{\rm{\;}}\\{\psi _{ab}}\left( \theta  \right) = {{\cal L}_{ab}}\eta \left( \theta  \right)\in {\mathbb{R}^{{n_\theta }}},
\end{array}
\end{equation}
where $\mathcal{L}_{ab} \in \mathbb{R}^{n_{\theta} \times n}$ is a matrix to take handpicked "good" elements from $\eta(\theta)$.

So the existence of inverse mapping $\theta={\cal{F}}\left(\psi_{ab}\right)$ ensures a theoretical possibility to recalculate the parameters $\eta \left( \theta  \right)$  into $\kappa \left( \theta  \right){\rm{,\;}}{T_I}\left( \theta  \right)$. In subsection {\emph{A}} a method is proposed to obtain the regression equations w.r.t. $\kappa \left( \theta  \right){\rm{,\;}}{T_I}\left( \theta  \right)$ and ${x_{\delta 0}}$ from \eqref{eq12}. In subsection {\emph{B}} an adaptive observer is proposed that uses the above-mentioned regression equations and allows one to achieve the stated goal \eqref{eq5}.

{\bf{Remark 1.}} \emph{It should be noted that in the general case $\kappa \left( \theta  \right){\rm{,\;}}{T_I}\left( \theta  \right)$ do not satisfy the Lipschitz condition, so the problem \eqref{eq5} cannot be reduced to $\theta$ identification (the mappings $\kappa \left( {\hat \theta } \right){\rm{,\;}}{T_I}\left( {\hat \theta } \right)$ can become singular in the course of a transient process). The below-proposed approach overcomes this problem by a linear transformation and allows one to identify $\kappa \left( \theta  \right){\rm{,\;}}{T_I}\left( \theta  \right)$ without singularity burden operations.}

{\bf{Remark 2.}} \emph{Integration in filters \eqref{eq14} starts from a sufficiently large time instant $t_{\epsilon}$ in order to avoid negative effect of a possibly large exponentially decaying term (see the proof of Lemma 1).}

\subsection{Parametrization of equations w.r.t. $\kappa \left( \theta  \right){\rm{,\;}}{T_I}\left( \theta  \right){\rm{,\;}}{x_{\delta 0}}$}

First of all, we put forward several hypotheses that: \emph{a}) the mapping ${\cal F}\left( {{\psi _{ab}}} \right)$ can be transformed into a linear regression equation w.r.t $\theta$, and \emph{b}) using the parametrization w.r.t $\theta$, the linear regression equations w.r.t. ${\psi _d}\left( \theta  \right){\rm{,\;}}{T_I}\left( \theta  \right)$ can be obtained.

{\bf{Hypothesis 1.}} \emph{There exist heterogeneous in the sense of \eqref{eq1} mappings ${\cal G}{\rm{:\;}}{\mathbb{R}^{{n_\theta }}} \to {\mathbb{R}^{{n_\theta } \times {n_\theta }}}$, ${\cal S}{\rm{:\;}}{\mathbb{R}^{{n_\theta }}} \to {\mathbb{R}^{{n_\theta }}}$  such that:}
\begin{equation}\label{eq18}
\begin{array}{c}
{\cal S}\left( {{\psi _{ab}}} \right) = {\cal G}\left( {{\psi _{ab}}} \right){\cal F}\left( {{\psi _{ab}}} \right) = {\cal G}\left( {{\psi _{ab}}} \right)\theta {\rm{,}}\\
\end{array}
\end{equation}
\vspace{-15pt}
\begin{gather*}
    {\Pi _\theta }\left( {\Delta} \right){\cal G}\left( {{\psi _{ab}}} \right) = {{\cal T}_{\cal G}}\left( {{\Xi _{\cal G}}\left( {\Delta} \right){\psi _{ab}}} \right){\rm{,}}\\
{\Pi _\theta }\left( {\Delta} \right){\cal S}\left( {{\psi _{ab}}} \right) = {{\cal T}_{\cal S}}\left( {{\Xi _{\cal S}}\left( {\Delta} \right){\psi _{ab}}} \right){\rm{,}}
\end{gather*}
\emph{where $\det \left\{ {{\Pi _\theta }\left( {\Delta} \right)} \right\} \ge {\Delta ^{{\ell _\theta }}}\left( t \right){\rm{,\;}}rank\left\{ {{\cal G}\left( {{\psi _{ab}}} \right)} \right\} = {n_\theta }{\rm{,\;}}\linebreak{\ell _\theta } \ge 1$, ${\Xi _{\cal G}}\left( {\Delta} \right) \in {\mathbb{R}^{{\Delta _{\cal G}} \times {n_\theta }}}$, ${\Xi _{\cal S}}\left( {\Delta} \right) \!\in\! {\mathbb{R}^{{\Delta _{\cal S}} \times {n_\theta }}}$, {${{\cal T}_{\cal G}}{\rm{:\;}}{\mathbb{R}^{{\Delta _{\cal G}}}} \mapsto \linebreak \mapsto {\mathbb{R}^{{n_\theta } \times {n_\theta }}}$, ${{\cal T}_{\cal S}}{\rm{:\;}}{\mathbb{R}^{{\Delta _{\cal S}}}} \mapsto {\mathbb{R}^{{n_\theta }}}$}, and all mappings are known.}

Having introduced the notation 
\begin{gather*}
\begin{array}{l}
{{\cal Y}_{ab}}\left( t \right){\rm{:}}= {{\cal L}_{ab}}{\cal Y}\left( t \right)=\Delta\left(t\right)\psi_{ab}\left(\theta\right),\\{{\cal M}_\theta }\left( t \right){\rm{:}} = {\rm{det}}\left\{ {{{\cal T}_{\cal G}}\left( {{{\overline \Xi }_{\cal G}}\left( \Delta  \right){{\cal Y}_{ab}}} \right)} \right\},
\end{array}
\end{gather*}
the hypothesis is put forward that \eqref{eq18} can be transformed into regression equations w.r.t ${\psi _d}\left( \theta  \right){\rm{,\;}}{T_I}\left( \theta  \right)$.

{\bf{Hypothesis 2.}} \emph{There exist heterogenous in the sense of \eqref{eq1} mappings ${\cal Q}{\rm{:\;}}{\mathbb{R}^{{n_\theta }}} \to {\mathbb{R}^{n \times n}}{\rm{,\;}}{\cal P}{\rm{:\;}}{\mathbb{R}^{{n_\theta }}} \to {\mathbb{R}^{n \times n}}$ such that:}
\begin{equation}\label{eq19}
{\cal Q}\left( \theta  \right) = {\cal P}\left( \theta  \right){{T_{I}}}\left( \theta  \right){\rm{,}}
\end{equation}
\vspace{-20pt}
\begin{gather*}
\begin{array}{c}
{\Pi _{{T_{I}}}}\left( {{{\cal M}_\theta }} \right){\cal P}\left( \theta  \right) = {{\cal T}_{\cal P}}\left( {{\Xi _{\cal P}}\left( {{{\cal M}_\theta }} \right)\theta } \right){\rm{,}}\\
{\Pi _{{T_{I}}}}\left( {{{\cal M}_\theta }} \right){\cal Q}\left( \theta  \right) = {{\cal T}_{\cal Q}}\left( {{\Xi _{\cal Q}}\left( {{{\cal M}_\theta }} \right)\theta } \right){\rm{,}}
\end{array}
\end{gather*}
\emph{where $\det \left\{ {{\Pi _{{{T_{I}}}}}\left( {{{\cal M}_\theta }} \right)} \right\} \ge {\cal M}_\theta ^{{\ell _{{T_{I}}}}}\left( t \right),\;rank\left\{ {{\cal P}\left( \theta  \right)} \right\} = n{\rm{,}}\linebreak{\ell _{{T_{I}}}} \ge 1{\rm{,\;}}$${\Xi _{\cal Q}}\left( {{{\cal M}_\theta }} \right) \in {\mathbb{R}^{{\Delta _{\cal Q}} \times {n_\theta }}}$, ${\Xi _{\cal P}}\left( {{{\cal M}_\theta }} \right) \in {\mathbb{R}^{{\Delta _{\cal P}} \times {n_\theta }}}$, ${{\cal T}_{\cal P}}{\rm{:\;}}{\mathbb{R}^{{\Delta _{\cal P}}}} \mapsto {\mathbb{R}^{n \times n}}$, ${{\cal T}_{\cal Q}}{\rm{:\;}}{\mathbb{R}^{{\Delta _{\cal Q}}}} \mapsto {\mathbb{R}^{n \times n}}$ and all mappings are known.}

{\bf{Hypothesis 3.}} \emph{There exist heterogeneous in the sense of \eqref{eq1} mappings ${\cal W}{\rm{:\;}}{\mathbb{R}^{{n_\theta }}} \to {\mathbb{R}^n}{\rm{,\;}}{\cal R}{\rm{:\;}}{\mathbb{R}^{{n_\theta }}} \to {\mathbb{R}^{n \times n}}$ such that:}
\begin{equation}\label{eq20}
{\cal W}\left( \theta  \right) = {\cal R}\left( \theta  \right){\psi _d}\left( \theta  \right){\rm{,}}
\end{equation}
\vspace{-20pt}
\begin{gather*}
\begin{array}{c}
{\Pi _{{\psi _d}}}\left( {{{\cal M}_\theta }} \right){\cal R}\left( \theta  \right) = {{\cal T}_{\cal R}}\left( {{\Xi _{\cal R}}\left( {{{\cal M}_\theta }} \right)\theta } \right){\rm{,}}\\
{\Pi _{{\psi _d}}}\left( {{{\cal M}_\theta }} \right){\cal W}\left( \theta  \right) = {{\cal T}_{\cal W}}\left( {{\Xi _{\cal W}}\left( {{{\cal M}_\theta }} \right)\theta } \right){\rm{,}}
\end{array}
\end{gather*}
\emph{where ${\rm{det}}\left\{ {{\Pi _{{\psi _d}}}\left( {{{\cal M}_\theta }} \right)} \right\} \!\!\ge\!\! {\cal M}_\theta ^{{\ell _{{\psi _d}}}}\left( t \right){\rm{,\;}}rank\left\{ {{\cal R}\left( \theta  \right)} \right\} \!=\! n{\rm{,\;}}\linebreak{\ell _{{\psi _d}}} \!\ge\! 1$, ${\Xi _{\cal W}}\left( {{{\cal M}_\theta }} \right) \in {\mathbb{R}^{{\Delta _{\cal W}} \times {n_\theta }}}$, ${\Xi _{\cal R}}\left( {{{\cal M}_\theta }} \right) \in {\mathbb{R}^{{\Delta _{\cal R}} \times {n_\theta }}}$, ${{\cal T}_{\cal R}}{\rm{:\;}}{\mathbb{R}^{{\Delta _{\cal R}}}} \to {\mathbb{R}^{n \times n}}$, ${{\cal T}_{\cal W}}{\rm{:\;}}{\mathbb{R}^{{\Delta _{\cal W}}}} \to {\mathbb{R}^n}$ and all mappings are known.}

As it is thoroughly discussed in \cite{b16, b17, b18}, the hypotheses \eqref{eq18}-\eqref{eq20} are not restrictive and hold in case if the mappings ${\cal F}\left( {{\psi _{ab}}} \right){\rm{,\;}}{\psi _d}\left( \theta  \right){\rm{,\;}}{T_I}\left( \theta  \right)$ are polynomials w.r.t. $\theta$, which is quite common situation as far as practical scenarios are concerned. Particularly, for the physical systems described by mathematical models, which are obtained by methods of mathematical physics. The fact that \eqref{eq18}-\eqref{eq20} are met and the property ${\Xi _{\left( . \right)}}\left( {\omega} \right) = {\overline \Xi _{\left( . \right)}}\left( {\omega} \right)\omega \left( t \right)$ of heterogenous mappings \eqref{eq1} allow one to transform the regression equation \eqref{eq12} w.r.t. $\eta \left( \theta  \right)$ into the linear regression equation w.r.t. $\kappa \left( \theta  \right){\rm{,\;}}{T_I}\left( \theta  \right){\rm{,\;}}{x_{\delta 0}}$.

\textbf{Lemma 2.} \emph{The unknown parameters $\kappa \left( \theta  \right)$, ${T_I}\left( \theta  \right)$ and ${x_{\delta 0}}$ for all $t \ge {t_\epsilon}$ satisfy measurable regression equations:}
\begin{equation}\label{eq21}
\begin{array}{c}
{{\cal Y}_\kappa }\left( t \right) = {{\cal M}_\kappa }\left( t \right)\kappa \left( \theta  \right){\rm{,}}\\
\end{array}
\end{equation}
\vspace{-20pt}
\begin{gather*}
\begin{array}{c}
    {{\cal Y}_\kappa }\left( t \right) = {\rm{adj}}\left\{ {{\rm{blkdiag}}\left\{ {\Delta \left( t \right){I_{2n}}{\rm{,\;}}{{\cal M}_{{\psi _d}}}\left( t \right){I_n}} \right\}} \right\}{\begin{bmatrix}
{{\cal Y}\left( t \right)}\\
{{{\cal Y}_{{\psi _d}}}\left( t \right)}
\end{bmatrix}}{\rm{,}}\\
{{\cal M}_\kappa }\left( t \right) = {\rm{det}}\left\{ {{\rm{blkdiag}}\left\{ {\Delta \left( t \right){I_{2n}}{\rm{,\;}}{{\cal M}_{{\psi _d}}}\left( t \right){I_n}} \right\}} \right\}{\rm{,}}
\end{array}
\end{gather*}
\begin{equation}\label{eq22}
\begin{array}{c}
{{\cal Y}_{{T_I}}}\left( t \right) = {{\cal M}_{{T_I}}}\left( t \right){T_I}\left( \theta  \right),\\
\end{array}
\end{equation}
\vspace{-20pt}
\begin{gather*}
\begin{array}{c}
{{\cal Y}_{{T_I}}}\left( t \right) = {\rm{adj}}\left\{ {{{\cal T}_{\cal P}}\left( {{{\overline \Xi }_{\cal P}}\left( {{{\cal M}_\theta }} \right){{\cal Y}_\theta }} \right)} \right\}{{\cal T}_{\cal Q}}\left( {{{\overline \Xi }_{\cal Q}}\left( {{{\cal M}_\theta }} \right){{\cal Y}_\theta }} \right){\rm{,}}\\
{{\cal M}_{{T_I}}}\left( t \right) = {\rm{det}}\left\{ {{{\cal T}_{\cal P}}\left( {{{\overline \Xi }_{\cal P}}\left( {{{\cal M}_\theta }} \right){{\cal Y}_\theta }} \right)} \right\}{\rm{,}}
\end{array}
\end{gather*}
\begin{equation}\label{eq23}
\begin{array}{c}
{{\cal Y}_{{x_{\delta 0}}}}\left( t \right) = {{\cal M}_{{x_{\delta 0}}}}\left( t \right){x_{\delta 0}}{\rm{,}}\\
\end{array}
\end{equation}
\vspace{-20pt}
\begin{gather*}
 {{\cal Y}_{{x_{\delta 0}}}}\left( t \right) = {\rm{adj}}\left\{ {{V_f}\left( t \right)} \right\}{p_f}\left( t \right){\rm{,\;}}{{\cal M}_{{x_{\delta 0}}}}\left( t \right) = {\rm{det}}\left\{ {{V_f}\left( t \right)} \right\}{\rm{,}}   
\end{gather*}
\emph{where}

1) \emph{the regression ${{\cal Y}_\theta }\left( t \right) = {{\cal M}_\theta }\left( t \right)\theta$ is formed using the following equations:}
\begin{gather*}
\begin{array}{c}
{{\cal Y}_\theta }\left( t \right) = {\rm{adj}}\left\{ {{{\cal T}_{\cal G}}\left( {{{\overline \Xi }_{\cal G}}\left( \Delta  \right){{\cal Y}_{ab}}} \right)} \right\}{{\cal T}_{\cal S}}\left( {{{\overline \Xi }_{\cal S}}\left( \Delta  \right){{\cal Y}_{ab}}} \right){\rm{,}}\\
{{\cal M}_\theta }\left( t \right) = {\rm{det}}\left\{ {{{\cal T}_{\cal G}}\left( {{{\overline \Xi }_{\cal G}}\left( \Delta  \right){{\cal Y}_{ab}}} \right)} \right\}{\rm{,\;}}{{\cal Y}_{ab}}\left( t \right) = {{\cal L}_{ab}}{\cal Y}\left( t \right){\rm{,}}
\end{array}  
\end{gather*}

2) \emph{the regression ${{\cal Y}_{{\psi _d}}}\left( t \right) = {{\cal M}_{{\psi _d}}}\left( t \right){\psi _d}\left( \theta  \right)$ is formed using the following equations:}
\begin{gather*}
\begin{array}{c}
{{\cal Y}_{{\psi _d}}}\left( t \right) = {\rm{adj}}\left\{ {{{\cal T}_{\cal R}}\left( {{{\overline \Xi }_{\cal R}}\left( {{{\cal M}_\theta }} \right){{\cal Y}_\theta }} \right)} \right\}{{\cal T}_{\cal W}}\left( {{{\overline \Xi }_{\cal W}}\left( {{{\cal M}_\theta }} \right){{\cal Y}_\theta }} \right){\rm{,}}\\
{{\cal M}_{{\psi _d}}}\left( t \right) = {\rm{det}}\left\{ {{{\cal T}_{\cal R}}\left( {{{\overline \Xi }_{\cal R}}\left( {{{\cal M}_\theta }} \right){{\cal Y}_\theta }} \right)} \right\},
\end{array}   
\end{gather*}

3) \emph{the signals ${p_f}\left( t \right)$ and ${V_f}\left( t \right)$ are obtained as follows:}
\begin{gather*}
\begin{array}{l}
{p_f}\left( t \right) = \int\limits_{{t_\epsilon}}^t {{e^{ -\sigma(\tau-t_{\epsilon})}}\left( {\Delta \left( \tau  \right){{\left( {{I_{{n_\delta }}} \otimes {{\cal Y}_{{\psi _d}}}\left( \tau  \right)} \right)}^{\rm{T}}}{V^{\rm{T}}}\left( \tau  \right) \times } \right.} \\
\end{array}    
\end{gather*}
\begin{equation}\label{eq24}
\begin{array}{l}
\left. { \times {C_0}{{\cal M}_{{\psi _d}}}\left( \tau  \right)p\left( \tau  \right)} \right)d\tau {\rm{,\;}}{p_f}\left( {{t_\epsilon}} \right) = {0_{{n_\delta }}}{\rm{,}}
\end{array}    
\end{equation}
\vspace{-15pt}
\begin{gather*}
\begin{array}{c}
p\left( t \right) = \Delta \left( t \right)\overline q\left( t \right) - C_0^{\rm{T}}\Omega \left( t \right){{\cal L}_a}{\cal Y}\left( t \right) - C_0^{\rm{T}}P\left( t \right){{\cal L}_b}{\cal Y}\left( t \right){\rm{,}}\\
{{\cal L}_a}{\psi _{ab}}\left( \theta  \right) = {\psi _a}\left( \theta  \right){\rm{,\;}}{{\cal L}_b}{\psi _{ab}}\left( \theta  \right) = {\psi _b}\left( \theta  \right),\\
{V_f}\left( t \right) \!=\! \int\limits_{{t_\epsilon}}^t {{\!e^{ -\sigma(\tau-t_{\epsilon})}}\left( {{\Delta ^2}\left( \tau  \right){{\left( {{I_{{n_\delta }}} \otimes {{\cal Y}_{{\psi _d}}}\left( \tau  \right)} \right)}^{\rm{T}}}{V^{\rm{T}}}\left( \tau  \right){C_0} \times } \right.} \\
 \times C_0^{\rm{T}}V\left( \tau  \right)\left( {{I_{{n_\delta }}} \otimes {{\cal Y}_{{\psi _d}}}\left( \tau  \right)} \right)d\tau {\rm{,\;}}{V_f}\left( {{t_\epsilon}} \right) = {0_{{n_\delta } \times {n_\delta }}},\\
\dot V\left( t \right) = {A_K}V\left( t \right) + \left( {h_\delta ^{\rm{T}}{\Phi _\delta }\left( t \right) \otimes {I_n}} \right){\rm{,\;}}V\left( {{t_0}} \right) = {0_{n \times n{n_\delta }}}{\rm{,}}
\end{array}    
\end{gather*}
\emph{and, if $\overline \varphi  \in {\rm{FE}}$ and $\left( {h_\delta ^{\rm{T}}{\Phi _\delta } \otimes {I_n}} \right) \in {\rm{FE}}$ over $\left[ {{t_\epsilon}{\rm{;\;}}{t_e}} \right]$, then for all $t \ge {t_e}$ it holds that:}
\begin{gather*}
\left| {{{\cal M}_\kappa }\left( t \right)} \right| \ge \underline {{{\cal M}_\kappa }}  > 0{\rm{,\;}}\left| {{{\cal M}_{{T_I}}}\left( t \right)} \right| \ge \underline {{{\cal M}_{{T_I}}}}  > 0{\rm{,\;}}\\\left| {{{\cal M}_{{x_{\delta 0}}}}\left( t \right)} \right| \ge \underline {{{\cal M}_{{x_{\delta 0}}}}}  > 0.   
\end{gather*}

\emph{Proof of Lemma 2 is given in Supplementary material \cite{b20}.}

~

\textbf{Remark 3.} \emph{The elements of the matrices ${\cal G}\left( {{\psi _{ab}}} \right){\rm{,\;}}{\cal P}\left( \theta  \right){\rm{,\;}}{\cal R}\left( \theta  \right)$ are handpicked as the denominators of the mappings ${\cal F}\left( {{\psi _{ab}}} \right){\rm{,\;}}{T_I}\left( \theta  \right){\rm{,\;}}{\psi _d}\left( \theta  \right)$, respectively. In their turn, ${\cal S}\left( {{\psi _{ab}}} \right){\rm{,\;}}{\cal Q}\left( \theta  \right){\rm{,\;}}{\cal W}\left( \theta  \right)$  are obtained as the result of direct multiplication – see their definitions in \eqref{eq18}-\eqref{eq20}.}

\subsection{Adaptive Observer Design}

Based on the regression equations \eqref{eq21}-\eqref{eq23} obtained in the previous subsection and using parameterizations \eqref{eq9}-\eqref{eq13}, the estimates of states and external disturbances are obtained as:
\begin{equation}\label{eq25}
\begin{array}{l}
\hat x\left( t \right) = {{\hat T}_I}\left( t \right)\hat \xi \left( t \right) = {{\hat T}_I}\left( t \right)\left( {z\left( t \right) + {{\hat R}^{\rm{T}}}\left( t \right)\hat \kappa \left( t \right)} \right){\rm{,}}\\
\hat \delta \left( t \right) = h_\delta ^{\rm{T}}{\Phi _\delta }\left( t \right){{\hat x}_{\delta 0}}\left( t \right){\rm{,}}\\
\dot {\hat U}\left( t \right) = {A_K}\hat U\left( t \right) + {I_n}\hat \delta \left( t \right){\rm{,\;}}\hat{U}\left( {{t_0}} \right) = {0_{n \times n}}{\rm{,}}
\end{array}
\end{equation}
where ${\hat R^{\rm{T}}}\left( t \right) = {\begin{bmatrix}
{\Omega \left( t \right)}&{P\left( t \right)}&{\hat U\left( t \right)}
\end{bmatrix}}{\rm{,}}$ and estimates $\hat \kappa \left( t \right){\rm{,\;}}{\hat T_I}\left( t \right){\rm{,\;}}{\hat x_{\delta 0}}\left( t \right)$ are obtained with the help of the following differential equations:
\begin{gather*}
\begin{array}{l}
\dot {\hat \kappa} \left( t \right) = \dot {\tilde \kappa} \left( t \right) =  - {\gamma _\kappa }{{\cal M}_\kappa }\left( t \right)\left( {{{\cal M}_\kappa }\left( t \right)\hat \kappa \left( t \right) - {{\cal Y}_\kappa }\left( t \right)} \right),\\
\end{array}
\end{gather*}
\vspace{-15pt}
\begin{equation}\label{eq26}
\begin{array}{l}
{{\dot {\hat x}}_{\delta 0}}\left( t \right) = {{\dot {\tilde x}}_{\delta 0}}\left( t \right) = \\ =  - {\gamma _{{x_{\delta 0}}}}{{\cal M}_{{x_{\delta 0}}}}\left( t \right)\left( {{{\cal M}_{{x_{\delta 0}}}}\left( t \right){{\hat x}_{\delta 0}}\left( t \right) - {{\cal Y}_{{x_{\delta 0}}}}\left( t \right)} \right),
\end{array}
\end{equation}
\vspace{-15pt}
\begin{gather*}
\begin{array}{l}
{{\dot {\hat T}_I}}\left( t \right) = {{\dot {\tilde T}_I}}\left( t \right) =  - {\gamma _{{T_I}}}{{\cal M}_{{T_I}}}\left( t \right)\left( {{{\cal M}_{{T_I}}}\left( t \right){{\hat T}_I} - {{\cal Y}_{{T_I}}}\left( t \right)} \right), \\
{\gamma _\kappa } > 0,{\rm{\;}}{\gamma _{{x_{\delta 0}}}} > 0,{\rm{\;}}{\gamma _{{T_I}}} > 0,
\end{array}
\end{gather*}
where $\tilde \kappa \left( t \right) = \hat \kappa \left( t \right) - \kappa \left( \theta  \right){\rm{,\;}}{\tilde x_{\delta 0}}\left( t \right) = {\hat x_{\delta 0}}\left( t \right) - {x_{\delta 0}}{\rm{,\;}}{\tilde T_I}\left( t \right) = \linebreak = {\hat T_I}\left( t \right) - {T_I}\left( \theta  \right)$ are parametric errors.

With the stated goal \eqref{eq5} in mind, the properties of the adaptive observer \eqref{eq25}, \eqref{eq26} are analyzed in Theorem.

\textbf{Theorem.} \emph{Let Assumptions 1-3 and Hypotheses \eqref{eq18}-\eqref{eq20} be met, then, if $\overline \varphi \in {\rm{FE}}$ and $\left( {h_\delta ^{\rm{T}}{\Phi _\delta } \otimes {I_n}} \right) \in {\rm{FE}}$, the observer \eqref{eq25}, \eqref{eq26} ensures that the goal \eqref{eq5} is achieved and additionally guarantees that:}
\begin{equation}\label{eq27}
\begin{array}{c}
\mathop {{\rm{lim}}}\limits_{t \to \infty } \left\| {{{\tilde x}_{\delta 0}}\left( t \right)} \right\| = {\rm{0 }}\left( {\exp} \right){\rm{,\;}}\mathop {{\rm{lim}}}\limits_{t \to \infty } \left\| {\tilde \kappa \left( t \right)} \right\| = {\rm{0}}\left( {\exp} \right){\rm{,}}\\
\mathop {{\rm{lim}}}\limits_{t \to \infty } \left\| {{{\tilde T}_I}\left( t \right)} \right\| = {\rm{0\;}}\left( {\exp} \right){\rm{,\;}}\\\mathop {{\rm{lim}}}\limits_{t \to \infty } \left\| {\hat U\left( t \right) - U\left( t \right)} \right\| = \mathop {{\rm{lim}}}\limits_{t \to \infty } \left\| {\tilde U\left( t \right)} \right\| = {\rm{0\;}}\left( {\exp} \right).
\end{array}
\end{equation}

\emph{Proof of Theorem is given in Supplementary material \cite{b20}.}

~

Thus, to solve the problem \eqref{eq5}, first of all, using the signals measured from system \eqref{eq3} and transformations \eqref{eq14}-\eqref{eq16}, the regression equations \eqref{eq12}, \eqref{eq13} are formed w.r.t. the unknown parameters $\eta \left( \theta  \right)$ and states $\xi \left( t \right)$ of the observer canonical form \eqref{eq7}. After that, if conditions \eqref{eq17}-\eqref{eq20} of polynomial overparameterization of system \eqref{eq3} are met, then the obtained regression equation is transformed via \eqref{eq21}-\eqref{eq23} into linear regression equations w.r.t. similarity matrix ${T_I}\left( \theta  \right)$, parameters $\kappa \left( \theta  \right)$ and initial conditions ${x_{\delta 0}}$, respectively. Then, based on the obtained regression equations, the identification laws \eqref{eq26} are derived, and an algebraic state observer \eqref{eq25} is introduced. In accordance with the results of the theorem, the obtained observer guarantees that the goal \eqref{eq5} is achieved if weak conditions $\overline \varphi \in {\rm{FE}}$ and $\left( {h_\delta ^{\rm{T}}{\Phi _\delta } \otimes {I_n}} \right) \in \linebreak \in {\rm{FE}}$ are met. Moreover, the condition $\overline \varphi \in {\rm{FE}}$ can be: \linebreak
\emph{i}) verified online by checking that eigenvalues of $\varphi \left( t \right)$ are non-zero, \emph{ii}) met by dither signal injection, and the condition $\left( {h_\delta ^{\rm{T}}{\Phi _\delta } \otimes {I_n}} \right) \in {\rm{FE}}$ can be verified offline because signal $h_\delta ^{\rm{T}}{\Phi _\delta }\left( t \right)$ is \emph{a priori} known according to Assumption 2.

\section{Numerical Experiments}
The system used for experimetns in \cite{b18} was considered:
\begin{equation}\label{eq28}
\begin{gathered}
  \dot x = {\begin{bmatrix}
  0&{{\theta _1} + {\theta _2}}&0 \\ 
  { - {\theta _2}}&0&{{\theta _2}} \\ 
  0&{ - {\theta _3}}&0 
\end{bmatrix}} x + {\begin{bmatrix}
  0 \\ 
  0 \\ 
  {{\theta _3}} 
\end{bmatrix}} u+ {\begin{bmatrix}
{{\theta _1}{\theta _2}}\\
0\\
0
\end{bmatrix}}\delta, \hfill \\
  y = {\begin{bmatrix}
  0&0&1 
\end{bmatrix}} x. \hfill \\ 
\end{gathered}
\end{equation}

The exosystem \eqref{eq4} parameters were set as:
\begin{gather*}
{{\cal A}_\delta } = \begin{bmatrix}
0&1\\
{ - 10}&{ 0}
\end{bmatrix}{\rm{,\;}}h_\delta ^{\rm{T}} = {\begin{bmatrix}
1&0
\end{bmatrix}}.   
\end{gather*}

Due to the space limitation, the particular forms of $\psi_a,\; \psi_b,\; \psi_d$, and all mappings are given in \cite{b20}.

The control signal was defined as a P-controller \linebreak $u =  - 75\left( {r - y} \right)$. The reference signal $r$ and parameters of the system \eqref{eq28} were chosen as:
\begin{equation}\label{eq29}
\begin{gathered}
r = 100 + 2{\text{.5}}{e^{ - \int\limits_{t_\epsilon}^{{t}} {1 d\tau} }}{\text{sin}}\left( {10t} \right){\text{, }}\\
{\theta _1} = {\theta _2} = 1{\text{, }}{\theta _3} =  - 1.
\end{gathered}
\end{equation}

The initial conditions of the system \eqref{eq28}, \eqref{eq4}, parameters of filters \eqref{eq14}-\eqref{eq16}, \eqref{eq24} and laws \eqref{eq26} were set as:
\begin{gather*}
\begin{array}{c}
K = { {\begin{bmatrix}
3&3&1
\end{bmatrix}}^{\rm{T}}}{\rm{,\;}}G = {\begin{bmatrix}
{ - 4}&1\\
{ - 2}&0
\end{bmatrix}}{\rm{,\;}}l = {\begin{bmatrix}
1\\
2
\end{bmatrix}}{\rm{,\;}}\beta  = {\begin{bmatrix}
{20}\\
{ - 8}
\end{bmatrix}}{\rm{,}}\\
{x_0} = {{\begin{bmatrix}
{ - 1}&0&2
\end{bmatrix}}^{\rm{T}}}{\rm{,\;}}{x_{\delta 0}} = {{\begin{bmatrix}
{500}&{100}
\end{bmatrix}}^{\rm{T}}}{\rm{,\;}}\\\hat \kappa \left( 0 \right) = {{\rm{0}}_9}{\rm{,\;}}{{\hat T}_I}\left( 0 \right) = {0_{3 \times 3}}{\rm{,}}\\
{{\hat x}_{\delta 0}}\left( 0 \right) = {0_2}{\rm{,\;}}{k_1} = 25,{\rm{\;}}k = {10^{19}}{\rm{,\;}}\sigma  = 1,{\rm{\;}}{t_\epsilon} = 25,{\rm{ }}\\
{\gamma _\kappa } = {10^{ - 74}}{\rm{,\;}}
{\gamma _{{x_{\delta 0}}}} = 5 \cdot {10^{ - 82}}
{\rm{,\;}}{\gamma _{{T_I}}} = {10^{ - 23}}.
\end{array}
\end{gather*}

The adaptive gains ${\gamma _\kappa }{\rm{,\;}}
{\gamma _{{x_{\delta 0}}}}{\rm{,\;}}{\gamma _{{T_I}}}$ had relatively small orders $10^{-74}$, $10^{-82}$ and $10^{-23}$, respectively, because the orders of regressors ${\cal{M}}_{\kappa}\left(t\right),\; {\cal{M}}_{x_{\delta0}}\left(t\right)$ and ${\cal{M}}_{T_{I}}\left(t\right)$ were relatively high in the course of simulations. In case we chose adaptive gains, for example, of the minus first order, then we would obtain stiff differential equations, which could not be solved numerically.
The parameters of the adaptive laws \eqref{eq26} were chosen so as to ensure rate of convergence, which coincided with the one for the adaptive laws in \cite{b18}.

Figure 1 for all $t \ge 25$ depicts transients of $\tilde x\left( t \right)$ and $\tilde \delta \left( t \right)$ for the observer \eqref{eq25} and the one from \cite{b18}.
\begin{figure}[htbp]
\centerline{\includegraphics[scale=0.55]{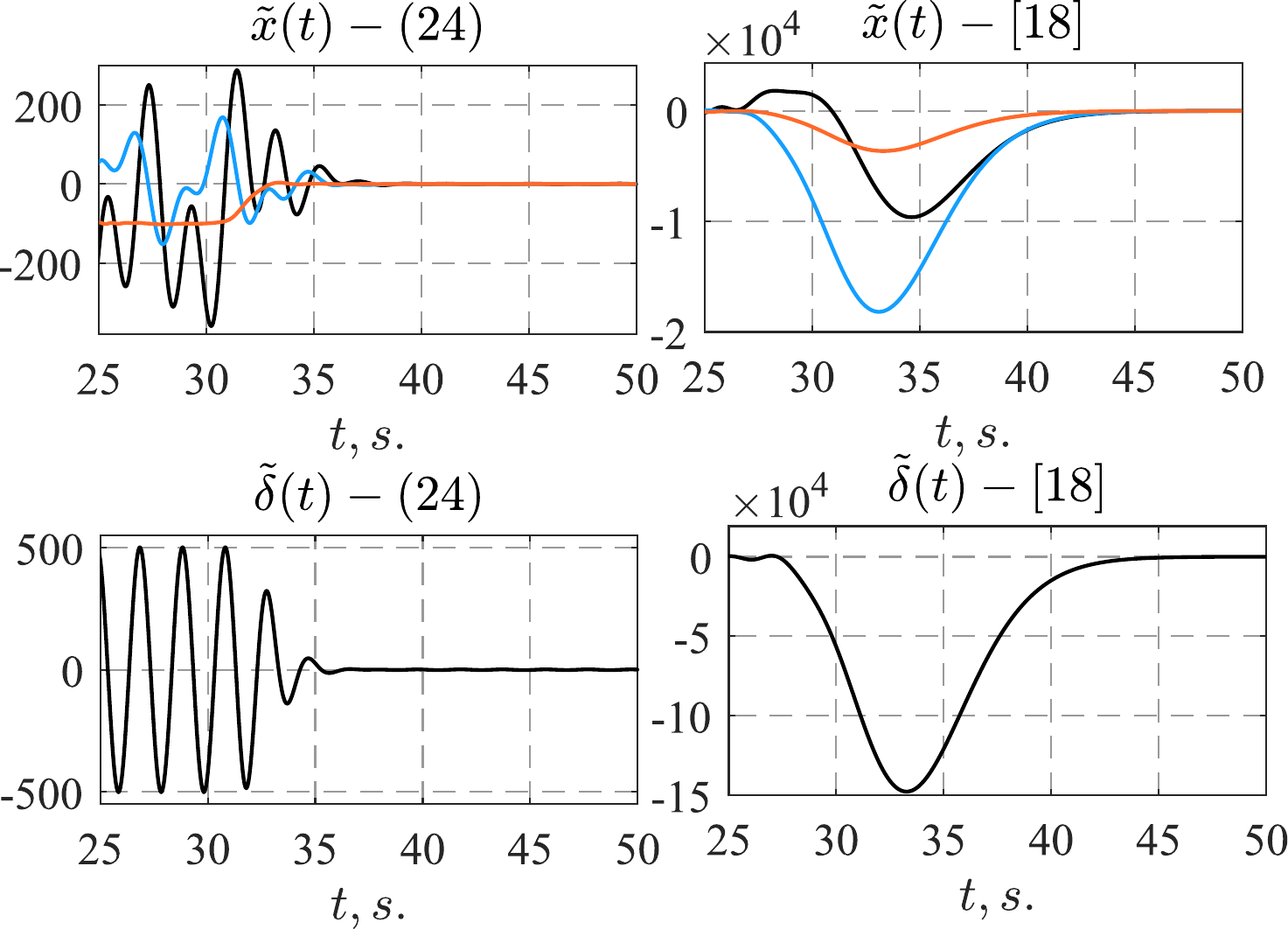}}
	\caption{Behavior of state $\tilde x\left( t \right)$ and disturbance $\tilde \delta \left( t \right)$ observation errors}
	\label{fig1}
\end{figure}

For all $t \ge 25$ Figure 2 presents transients of parametric errors $\tilde \kappa \left( t \right){\rm{,\;}}{\tilde x_{\delta 0}}\left( t \right){\rm{,\;}}{\tilde T_I}\left( t \right)$ and the norm $\left\| {\tilde U\left( t \right)} \right\|$.

\begin{figure}[htbp]
\centerline{\includegraphics[scale=0.55]{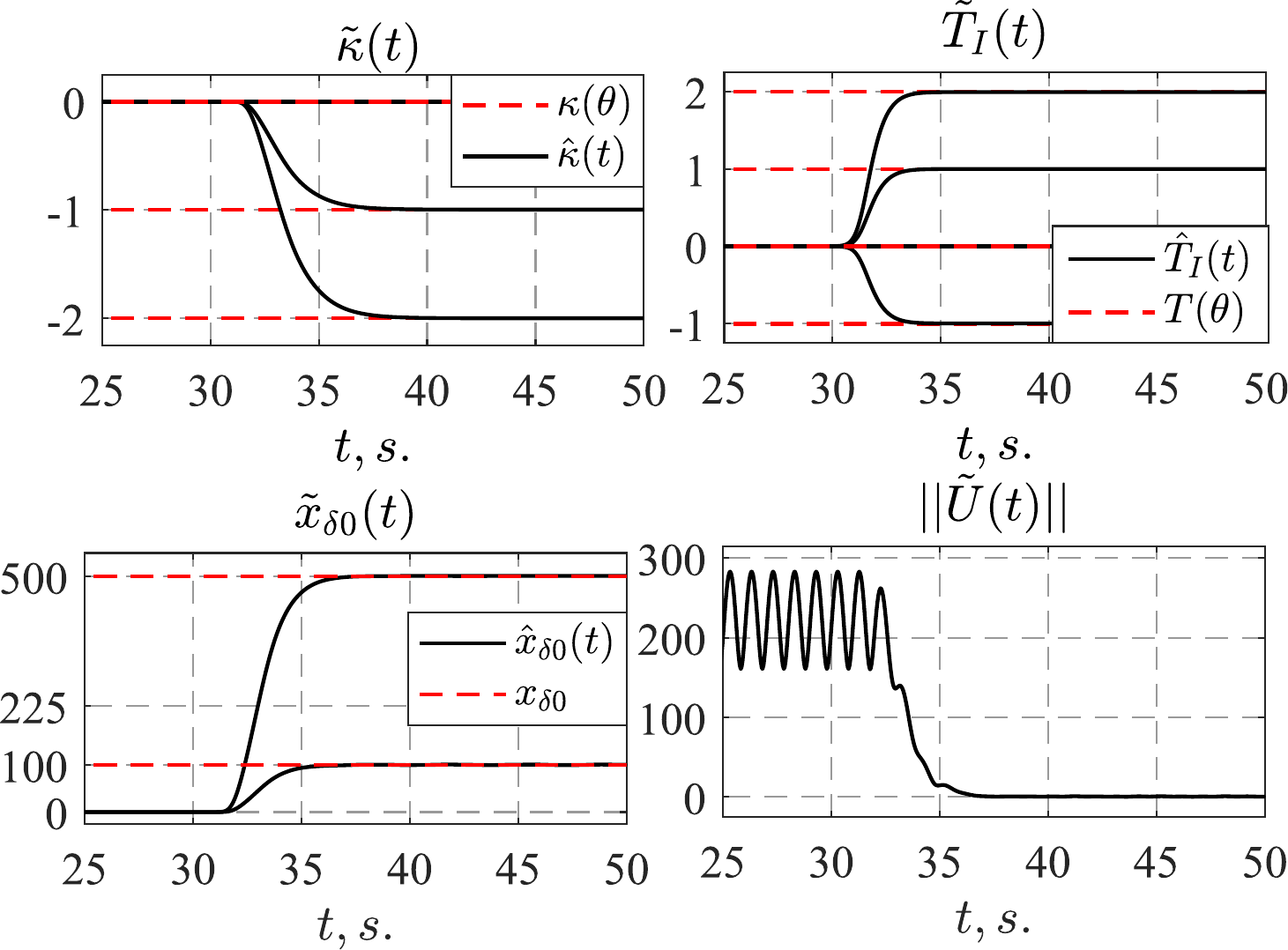}}
	\caption{Behavior of $\left\| {\tilde U\left( t \right)} \right\|$ and parametric errors $\tilde \kappa \left( t \right){\rm{,\;}}{\tilde x_{\delta 0}}\left( t \right){\rm{,\;}}{\tilde T_I}\left( t \right)$}
	\label{fig2}
\end{figure}

Therefore, when the parameter identification process had been completed, the unmeasured state and disturbance observation errors exponentially converged to zero. As the estimates were obtained using the algebraic equations \eqref{eq25}, then, in comparison with the observer from \cite{b18}, there was no significant overshoot in the course of the transients.

\section{Conclusion}

For the uncertain linear time-invariant overparametrized systems an extended adaptive observer was proposed that, if the regressor is finitely exciting, allowed one to reconstruct unmeasured states and bounded external disturbance.

In contrast to solutions \cite{b2, b3, b8, b10, b13, b14}, the proposed observer: \emph{i}) allowed one to reconstruct physical states $x\left( t \right)$ of the system \eqref{eq3} rather than virtual ones $\xi \left( t \right)$ of the observer canonical form \eqref{eq7}, \emph{ii}) provided exponential convergence of the unmeasured state and disturbance observation errors if the regressor finite excitation condition was met. In contrast to the earlier result \cite{b18}, the proposed adaptive observer: (\emph{a}) formed unmeasured states and external perturbation estimates using the algebraic equation, and (\emph{b}) was not affected by the peaking phenomenon (significant overshoot in the course of transients of $\tilde x\left( t \right)$ and $\tilde \delta \left( t \right)$).


\begin{thebibliography}{00}
\bibitem{b1} Bernard P., Andrieu V., Astolfi D., ``Observer design for continuous-time dynamical systems,'' Annual Reviews in Control, vol. 53, pp. 224--248, 2022.
\bibitem{b2} Carroll R., Lindorff D., ``An adaptive observer for single-input single-output linear systems,'' IEEE Transactions on Automatic Control, vol. 18, no. 5, pp. 428--435, 1973.
\bibitem{b3} Luders G., Narendra K. S., ``An adaptive observer and identifier for a linear system,'' IEEE Transactions on Automatic Control, vol. 18, no. 5, pp. 496--499, 1973.
\bibitem{b4} Zhang Q., Delyon B., ``A new approach to adaptive observer design for MIMO systems,'' Proceedings of the 2001 American Control Conference, vol. 2, pp. 1545--1550, 2001.
\bibitem{b5} Kazantzis N., Kravaris C., ``Nonlinear observer design using Lyapunov’s auxiliary theorem,'' Systems $\&$ Control Letters, vol. 34, no. 5, pp. 241--247, 1998.
\bibitem{b6} Ortega R., et al., ``Generalized parameter estimation-based observers: Application to power systems and chemical–biological reactors,'' Automatica, vol. 129, pp. 109635, 2021.
\bibitem{b7} Bobtsov A., Ortega R., Yi B., Nikolaev N., ``Adaptive state estimation of state-affine systems with unknown time-varying parameters,'' International Journal of Control, vol. 95, no. 9, pp. 2460--2472, 2022.
\bibitem{b8} Zhang Q., ``Adaptive observer for multiple-input-multiple-output (MIMO) linear time varying systems,'' IEEE Transactions on Automatic Control, vol. 47, no. 3, pp. 525--529, 2002.
\bibitem{b9} Zhang Q., ``Revisiting different adaptive observers through a unified formulation,'' in Proc. 44th Conference on Decision and Control, pp. 3067--3072, 2005.
\bibitem{b10} Kreisselmeier G., ``Adaptive observers with exponential rate of convergence,'' IEEE Trans. on Automatic Control, vol. 22, pp. 2--8, 1977.
\bibitem{b11} Marino R., Tomei P., ``Adaptive observers with arbitrary exponential rate of convergence for nonlinear systems,'' IEEE Transactions on Automatic Control, vol. 40, no. 7, pp. 1300--1304, 1995.
\bibitem{b12} Marino R., Santosuosso G. L., Tomei P., ``Robust adaptive observers for nonlinear systems with bounded disturbances,'' IEEE Transactions on Automatic Control, vol. 46, no. 6, pp. 967--972, 2001.
\bibitem{b13} Katiyar A., Roy S. B., Bhasin S., ``Initial Excitation Based Robust Adaptive Observer for MIMO LTI Systems,'' IEEE Transactions on Automatic Control, vol. 68, no.4, pp. 2536-2543, 2023.
\bibitem{b14} Bobtsov A., Pyrkin A., Vedyakov A., Vediakova A., Aranovskiy S. ``A Modification of Generalized Parameter-Based Adaptive Observer for Linear Systems with Relaxed Excitation Conditions,'' IFACPapersOnLine. vol. 55, no. 12., pp.324--329, 2022.
\bibitem{b15} Szabat K., Orlowska-Kowalska T., ``Vibration suppression in a two-mass drive system using PI speed controller and additional feedbacks – Comparative study,'' IEEE Transactions on Industrial Electronics, vol. 54, no. 2., pp.1193--1206, 2007.
\bibitem{b16} Glushchenko A., Lastochkin K., ``Exponentially Stable Adaptive Observation for Systems Parameterized by Unknown Physical Parameters,'' arXiv preprint arXiv:2212.08405. pp.1--8, 2023.
\bibitem{b17} Glushchenko A., Lastochkin K., ``Parameter Estimation-Based Observer for Linear Systems with Polynomial Overparameterization,'' in Proc. 31st Mediterranean Conference on Control and Automation (MED), pp.7955-799, 2023.
\bibitem{b18} Glushchenko A., Lastochkin K., ``Extended Adaptive Observer for Linear Systems with Overparameterization,'' in Proc. 31st Mediterranean Conference on Control and Automation (MED), pp. 789-794, 2023.
\bibitem{b100} Wang, L., Ortega, R., Bobtsov, A., Romero, G.J., ``Identifiability Implies Robust, Globally Exponentially Convergent On-line Parameter Estimation,'' International Journal of Control, 2023.
\bibitem{b19} Nikiforov V. O., ``Observers of external deterministic disturbances. II. Objects with unknown parameters,'' Automation and Remote Control, vol. 65, no.11, pp.1724--1732, 2004.
\bibitem{b20} Glushchenko A., Lastochkin K., ``Supplement to “Parameter Estimation-Based Extended Observer for Linear Systems with Polynomial Overparameterization,'' arXiv preprint arXiv:2302.13705. pp.1--6, 2023.{ \tiny \url{ https://arxiv.org/src/2302.13705v5/anc/supp.pdf}}.
\bibitem{b21} Walter E., Pronzato L. Identification of parametric models from experimental data. Berlin : Springer, 1997.
\end{thebibliography}
\end{document}